\documentclass[12pt,english]{article}
\usepackage{bm,bbm,lscape}
\usepackage{natbib, amsmath,authblk}
\usepackage{marvosym}
\usepackage{multicol}
\usepackage[british]{babel}
\usepackage{graphicx,xcolor,geometry,setspace,lineno}
\usepackage{pstricks}
\usepackage[british]{babel}
\usepackage[justification=justified,singlelinecheck=false,labelfont=it]{caption}
\usepackage{graphicx}
\usepackage{fancyhdr}
\definecolor{Mygrey}{gray}{0.65}

\title{Semiparametric stochastic volatility modelling using penalized splines}

  \author[1]{Roland Langrock}
  \author[2]{Th\'eo Michelot}
  \author[3]{Alexander Sohn}
  \author[3]{Thomas Kneib}

  \affil[1]{\small University of St Andrews, UK.}
  \affil[2]{\small INSA de Rouen, France}
  \affil[3]{\small Georg August University of G\"{o}ttingen, Germany}

\date{}

\begin{document}

\begin{spacing}{1.55}
\maketitle

\vspace{-1em}

\begin{abstract}
\noindent
Stochastic volatility (SV) models mimic many of the stylized facts attributed to time series of asset returns, while maintaining conceptual simplicity. The commonly made assumption of conditionally normally distributed or Student-t-distributed returns, given the volatility, has however been questioned. In this manuscript, we introduce a novel maximum penalized likelihood approach for estimating the conditional distribution in an SV model in a nonparametric way, thus avoiding any potentially critical assumptions on the shape. The considered framework exploits the strengths both of the powerful hidden Markov model machinery and of penalized B-splines, and constitutes a powerful and flexible alternative to recently developed Bayesian approaches to semiparametric SV modelling. We demonstrate the feasibility of the approach in a simulation study before outlining its potential in applications to three series of returns on stocks and one series of stock index returns.
\end{abstract}

\vspace{0em}

\section{Introduction}\label{intro}

Stochastic volatility (SV) models are immensely popular tools for the analysis of financial time series. This subclass of state-space models (SSMs) constitutes one of the two most widely used approaches
for modelling stock market volatility, the other one being ARCH/GARCH type models. The structure of the standard discrete-time SV model, labeled SV$_0$ in the following, for log-returns $y_1,\ldots,y_T$ on an asset, is as follows:
 \begin{linenomath*}
 \begin{equation}
 y_t= \varepsilon_t^{(0)} \beta \exp(g_t/2), \qquad  g_{t}= \phi g_{t-1}   + \sigma \eta_t, 
 \label{SVbasic}
 \end{equation}
 \end{linenomath*}
where $\beta, \sigma >0$, and where $ \{\varepsilon_t^{(0)}\}$ and $\{\eta_t\}$ are independent sequences of independent standard normal random variables (see, e.g., \citealp{she96}). Stationarity is obtained for $|\phi|<1$. The unobserved sequence $\{g_t\}$, commonly referred to as the log-volatility process, represents the time-varying ``nervousness'' of the market. This model captures several of the stylized facts attributed to asset returns, including positive autocorrelation of squared returns (indicating a volatility that slowly varies over time and hence volatility clustering), zero autocorrelation of the unsquared returns, and a kurtosis in excess of 3. However, the basic model tends to underestimate the probability of relatively extreme returns, such that it is often more adequate to consider a Student-t distribution with $\nu$ degrees of freedom for $\varepsilon_t^{(0)}$ \citep{chi02}; we label this second model SV$_{\text t}$.

In the existing literature, several different model formulations have been considered that extend the flexibility of the log-volatility process, $\{g_t\}$ (e.g., \citealp{gal97}; \citealp{abr06}; \citealp{lan12}). Yet \citet{dur06} found ``no evidence that even simple single  factor models are unable to capture the dynamics of the volatility process'' (p.276). Instead, he considers the shape of the conditional distribution in SV models --- i.e., of the conditional distribution of $y_t$, given $g_t$ --- to be ``the more critical problem'' (p.304). In addition to heavy tails, which are accounted for in the SV$_{\text t}$ formulation, evidence of asymmetries has been found (see, e.g., \citealp{gal97}; \citealp{har00}; \citealp{jon03}; \citealp{dur06}). To get the shape right, and in particular to accurately estimate the tails of the conditional distribution, is of high importance for example in risk management. While parametric models can be constructed that enable the inclusion of heavy tails and skewness, nonparametric approaches have the considerable advantage that no restriction to a particular class of distributions is made a priori.

Much recent work in this direction has been conducted in the Bayesian context, where the normal distribution can be used as a building block to formulate more complex models that still utilize the benefits of the normal formulation for constructing convenient update schemes in a Markov chain Monte Carlo simulation. \citet{aba10} consider scale mixtures of normals for the conditional distribution, where the variance of the normal distribution is supplemented with suitable prior specifications that yield a larger class of potential marginal distributions after integrating out the mixing distribution for the variance. Nonparametric specifications relying on an infinite-dimensional mixture of normals, generated by a Dirichlet process mixture prior, have been developed in \citet{jen10} and \citet{del11}. While \citet{jen10} directly tackle the conditional distribution of the returns, \citet{del11} employ a different representation where the logarithm of the squared noise in the return process is considered for the analysis. \citet{del13} extend the latter approach by including a leverage effect, allowing for the potential correlation of the two error terms, in the return and in the log-volatility process, respectively. 

In this manuscript, we develop a novel frequentist approach for nonparametrically estimating the conditional distribution in an SV model. The proposed maximum penalized likelihood approach exploits the strengths both of likelihood-based hidden Markov model (HMM) machinery and of penalized B-splines (i.e., P-splines). The former is employed to deal with a well-known difficulty with SV models, which is that their likelihood is given by a high-order multiple integral that is analytically intractable. It has however been shown that methods available for HMMs --- which have the same dependence structure as SV models and constitute another subclass of SSMs, with finite state space --- can be applied in order to perform a fast and accurate numerical integration of the SV model likelihood. More specifically, such a numerical integration corresponds to a fine discretization of the support of the log-volatility process. The associated transformation of the continuous support of $\{g_t\}$ to a finite support renders the powerful HMM forward algorithm applicable, making it feasible to evaluate an arbitrarily accurate approximation to the SV model likelihood \citep{fri98,bar01,bar03,lan12}. We extend this likelihood-based approach to allow for a nonparametric estimation of the conditional distribution, by representing the density of this distribution as a linear combination of a large number of standardized B-spline basis functions, including a roughness penalty in the likelihood in order to arrive at an appropriate balance between goodness of fit and smoothness for the fitted density. Since we still model the log-volatility process in a parametric way, we use the label SV$_{\text{sp}}$ to refer to the resulting {\em semiparametric} SV model with nonparametrically modelled conditional distribution. 

The paper is structured as follows. In Section \ref{model}, we begin by describing the HMM-based likelihood evaluation, then introducing the B-spline-based representation of the conditional distribution and discussing associated inferential issues. The performance of the suggested approach is investigated in a simulation study in Section \ref{simul}. In Section \ref{appl}, we apply the approach to real data related to three stocks and one stock index, comparing the predictive performance of our model to popular parametric counterparts.

\section{Semiparametric SV modelling}\label{model}

\subsection{SV model likelihood}\label{modelsubsec}

We consider a model SV$_{\text{sp}}$ of the form
 \begin{linenomath*}
\begin{equation}\label{SVnp}
y_t = \varepsilon_t \exp(g_t/2), \qquad g_{t} = \phi g_{t-1} + \sigma \eta_t , 
\end{equation}
 \end{linenomath*}
with the $\eta_t$ iid standard normal, but where, in contrast to the models SV$_0$ and SV$_{\text t}$, we do not make any assumptions on the distributional form of the random variables $\varepsilon_t$. However, we do assume these variables to be iid, and to be independent of $\{\eta_t\}$. Our aim is to nonparametrically estimate the probability density function (pdf) $f_{\varepsilon}$ of the variables $\varepsilon_t$. Compared to model SV$_0$, as given in (\ref{SVbasic}), we have omitted the parameter $\beta$ in (\ref{SVnp}), since otherwise the semiparametric model would not be identifiable; in the  SV$_{\text{sp}}$ model the effect of $\beta$ will be absorbed within $f_{\varepsilon}$. Before we introduce our strategy for estimating $f_{\varepsilon}$ in a nonparametric way (alongside the other model parameters), we will derive a tractable likelihood function for general $f_{\varepsilon}$, including the nonparametric case, but also those of a normal distribution and of a Student-t distribution. To formulate the likelihood, we will require the conditional pdfs of the random variables $y_t$, given $g_t$ ($t=1,\ldots,T$). We denote these conditional pdfs by $f(y_t|g_t)$, for $t=1,\ldots,T$. 
These pdfs are simple transformations of the density $f_{\varepsilon}$:
 \begin{linenomath*}
\begin{equation*}
f(y_t|g_t)     
  = \exp(-g_t/2) f_{\varepsilon}(y_t \exp(-g_t/2)) \, .
\end{equation*}
 \end{linenomath*}
For any $f_{\varepsilon}$, the likelihood of the model defined by (\ref{SVnp}) can then be derived as
 \begin{linenomath*}
\begin{align}\label{likl}
\nonumber \mathcal{L}  & = {\int \ldots \int} f(y_1,\ldots ,y_T,g_1,\ldots ,g_T) \, dg_T \ldots dg_1 \\
\nonumber              & = {\int \ldots \int} f(y_1,\ldots ,y_T | g_1,\ldots ,g_T) f(g_1,\ldots ,g_T) \, dg_T \ldots dg_1 \\
\nonumber              & = \int \ldots \int  f(g_1) f(y_1 \vert g_1) \prod_{t=2}^T f(g_t\vert g_{t-1})f(y_t\vert g_t) \, dg_T \ldots dg_1 \, \\
\nonumber              & = \int \ldots \int  f(g_1) \exp(-g_1/2)f_{\varepsilon}(y_1 \exp(-g_1/2)) \\
                       & \quad \times \prod_{t=2}^T f(g_t\vert g_{t-1}) \exp(-g_t/2)f_{\varepsilon}(y_t \exp(-g_t/2)) \, dg_T \ldots dg_1 \, .
\end{align}
 \end{linenomath*}
In the second last step, we exploited the dependence structure that is characteristic of SV models, HMMs and general SSMs. Hence, the likelihood is a high-order multiple integral that cannot be evaluated directly. Via numerical integration, using a simple rectangular rule based on $m$ equidistant intervals, $B_i = (b_{i-1},b_i)$, $i=1, \ldots , m$, with midpoints $b_i^*$ and of length $b$, the likelihood can be approximated as follows:
 \begin{linenomath*}
\begin{align}\label{approxL}
\nonumber \mathcal{L}  & \approx b^T \sum_{i_1=1}^m \ldots \sum_{i_T=1}^m f(b_{i_1}^*) \exp(- b_{i_1}^*/2) f_{\varepsilon} (y_1 \exp(-b_{i_1}^*/2)) \\
                       & \quad \times \prod_{t=2}^T f(b_{i_t}^* \vert b_{i_{t-1}}^*) \exp(- b_{i_t}^*/2)f_{\varepsilon}(y_t \exp(- b_{i_t}^*/2)) = \mathcal{L}_{\text{approx}} \, .
\end{align}
 \end{linenomath*}
This approximation can be made arbitrarily accurate by increasing $m$, and in fact virtually exact for  $m$ around 100, provided that the interval $(b_0,b_m)$ covers the essential range of the log-volatility process (more details below). In the given form, the approximate likelihood (\ref{approxL}) is usually computationally intractable, since it involves $m^T$ function evaluations. However, an efficient recursive scheme can be used to evaluate (\ref{approxL}). To see this, note that the numerical integration essentially corresponds to a discretization of the state space, i.e., of the support of the log-volatility process $\{ g_t\}$. Therefore, the approximate likelihood given in (\ref{approxL}) can be evaluated using the well-developed and powerful machinery of the subclass of SSMs given by HMMs, which are SSMs with a finite state space (cf.\ \citealp{lan11}; \citealp{lan12}). We sketch the relevant HMM methodology in the appendix to this manuscript. In particular, in the appendix we highlight a key property of HMMs, which is that the likelihood can be evaluated efficiently using the so-called forward algorithm \citep{zuc09}. 
For an HMM, applying the forward algorithm results in a matrix product expression for the likelihood, and this is exactly what we obtain also in the present context:
 \begin{linenomath*}
\begin{equation}\label{likelihood}
\mathcal{L}_{\text{approx}} = \boldsymbol{\delta} \mathbf{P}(y_1)\boldsymbol{\Omega}\mathbf{P}(y_2)\boldsymbol{\Omega}\mathbf{P}(y_3) \cdots  \boldsymbol{\Omega}\mathbf{P}(y_{T-1})\boldsymbol{\Omega}\mathbf{P}(y_T) \mathbf{1} \, .
\end{equation}
 \end{linenomath*}
Here, the $m \times m$-matrix $\boldsymbol{\Omega}=\bigl(\omega_{ij}\bigr)$ is the analogue to the transition probability matrix in case of an HMM (see the appendix), defined as $\omega_{ij}=  f(b_{j}^* \vert b_{i}^*) \cdot b$. Furthermore, the vector $\boldsymbol{\delta}$ is the analogue to the Markov chain initial distribution in case of an HMM, here defined such that ${\delta_i}$, $i=1,\ldots,m$, is the density of the normal distribution with mean zero and standard deviation $\sigma/\sqrt{1-\phi^2}$ --- the stationary distribution of the autoregressive process used to model the log-volatility ---  evaluated at $b_i^*$ and multiplied by $b$. Finally, $\mathbf{P}(y_t)$ is an $m \times m$ diagonal matrix with $i$th diagonal entry $\exp(- b_{i}^*/2)f_{\varepsilon}(y_t \exp(- b_{i}^*/2))$, hence the analogue to the matrix comprising the state-dependent probabilities in case of an HMM. Using the matrix product expression given in (\ref{likelihood}), the computational effort required to evaluate the approximate likelihood is linear in the number of observations, $T$, and quadratic in the number of intervals used in the discretization, $m$. In practice, this means that the likelihood of an SV model can typically be calculated in a fraction of a second, even for $T$ in the thousands and say $m=150$, a value which renders the approximation virtually exact. Furthermore, $\mathcal{L}_{\text{approx}} \rightarrow \mathcal{L}$ as $b_m, m \rightarrow \infty$ and $b_0 \rightarrow -\infty$.
 
\subsection{Nonparametric modelling}
\label{modelsubsec2}
 
We now turn to the nonparametric modelling of the distribution of $\varepsilon_t$. Following \citet{sch12}, we suggest to estimate the pdf of $\varepsilon_t$ by considering finite linear combinations of a large number of basis functions:
 \begin{linenomath*}
\begin{equation}\label{lincom}
\hat{f}_{\varepsilon}(x) = \sum_{k=-K}^K {a}_k \psi_k(x) \, .
\end{equation}
 \end{linenomath*}
Here the basis functions $\psi_{-K},\ldots,\psi_{K}$ are known and fixed pdfs. Clearly, $\hat{f}_{\varepsilon}(x)$ then is a pdf if $\sum_{k=-K}^K {a}_k =1$ and ${a}_j\geq 0$ for all $j=-K,\ldots,K$. To enforce these constraints, the coefficients to be estimated, $a_{-K},\ldots,a_{K}$, are transformed using the multinomial logit link function
 \begin{linenomath*}
\begin{equation}\label{ab}
a_k = \frac{\exp(\beta_k)}{\sum_{j=-K}^K \exp(\beta_j)} \, ,
\end{equation}
 \end{linenomath*}
where we set $\beta_0=0$ for identifiability. In principle, any set of densities $\psi_{-K},\ldots,\psi_{K}$ can be used to approximate $f_{\varepsilon}(x)$ as in (\ref{lincom}). We follow \citet{sch12} and use B-splines, in ascending order in the basis used in (\ref{lincom}), and standardized such that they integrate to 1. For more details on B-splines, see, e.g., \citet{deb78} and \citet{eil96}. Since each B-spline basis function is associated with a separate parameter, this model formulation in fact leads to a finite-dimensional parameter space. However, the dimensionality is high and the separate parameters are no longer of interest or interpretable --- the model is {\em overparameterized}. We therefore refer to our estimation approach as nonparametric despite the fact that it does rely on a parametric specification with a large number of parameters. This is in line with the standard terminology used in the literature, where (penalized) spline approaches are subsumed under nonparametric approaches \citep[see, e.g.,][]{rup03}.

The approximate likelihood of the resulting  SV$_{\text{sp}}$ model is given by (\ref{likelihood}), plugging in $\hat{f}_{\varepsilon}$ for ${f}_{\varepsilon}$ in the matrices $\mathbf{P}(y_t)$, $t=1,\ldots,T$. Following \citet{eil96}, we modify the (approximate) log-likelihood by including a penalty on ($q$-th order) differences between coefficients associated with adjacent B-splines, yielding the penalized log-likelihood
 \begin{linenomath*}
\begin{equation}\label{penlik}
l_p = \log \bigl( \mathcal{L}_{\text{approx}} \bigr) - \frac{\lambda}{2} \sum_{k=-K+q}^{K} \bigl( \Delta^q a_k \bigr)^2  \, ,
\end{equation}
 \end{linenomath*}
with $a_k$ parameterized as in (\ref{ab}) and smoothing parameter $\lambda \geq 0$. The penalty term involves the difference operator $\Delta$, where $\Delta a_k = a_k-a_{k-1}$ and $\Delta^q a_k = \Delta (\Delta^{q-1} a_k)$. This results in a penalization of roughness of the estimator, with $\lambda$ controlling how much emphasis is put on goodness of fit and on smoothness, respectively.
In particular, unpenalized estimates are obtained for ${\lambda}=0$. For $\lambda\rightarrow\infty$ the penalty will dominate the likelihood, resulting in a sequence of weights $a_{k}$ that follow a polynomial of order $q-1$ in $k$. The difference order therefore also affects the smoothness of the estimates indirectly (and to a much smaller extent than the degree of the spline basis). We will use $q=2$ in the remainder since this provides an approximation to the integrated squared second derivative penalty that is popular in the context of smoothing splines.

Including the penalty term in the likelihood avoids the problem of selecting an optimal number of basis elements, since the penalty effectively reduces the number of free basis parameters and yields an adaptive fit to the data, provided the smoothing parameter is chosen in a data-driven way. The number of basis elements needs to be large enough to give sufficient flexibility for reflecting the structure of the conditional distribution $f_{\varepsilon}$, but once this threshold is passed, increasing the number of basis elements further does no longer change the fit to the data much due to the impact of the penalty. For moderately smooth regression functions, \citet{rup02} recommends to use a default of about 35 (or 40) basis functions. To capture the pdf of $\varepsilon_t$ in an SV model, we expect such a choice to easily provide sufficient flexibility, and hence have chosen $K$ accordingly in our analyses (see below). To select the smoothing parameter in a data-driven way, we will consider cross-validation (see Section \ref{smoothchoice}).

In preliminary simulation experiments, we found that with an equidistant spacing of the knots our approach tended to produce estimated densities that were overly smooth around the peak of the true distribution and too wiggly in the tails. This is related to the fact that the basis coefficients systematically decay towards the tails of the estimated distribution which would require an adaptive amount of smoothing instead of a global smoothing parameter. As a simple yet effective strategy to achieve such adaptiveness, we consider increasingly wider distances between the B-spline basis densities towards the tails instead of the common equidistant specification. Since we still rely on the unweighted difference penalty in (\ref{penlik}), this effectively increases the penalty for the tails of the distribution.
 
\subsection{Inference}
\label{inference}
 
\subsubsection{Parameter estimation}
 
The use of the forward algorithm leads to a very fast evaluation of the penalized log-likelihood given in (\ref{penlik}). A numerical maximization of the penalized log-likelihood is therefore feasible in typical cases, even for high $m$ and hence very close approximations to the likelihood in (\ref{likl}); some computing times are given in Section \ref{appl}. Since the first part of expression (\ref{penlik}) is susceptible to numerical overflow, it is required to compute its logarithm, which involves a minor difficulty since we are dealing with a matrix product. However, techniques to address this issue are standard: \citet{zuc09} describe a straightforward scaling strategy for calculating the logarithm of an HMM-type matrix product likelihood (see their Chapter 3).

In practice, one also has to select the value of $m$, the number of intervals used in the discretization of the log-volatility process, and the range of possible $g_t$-values considered in the numerical integration. In our experience, estimates usually stabilize for values of $m$ around 50 (cf.\ \citealp{lan12}; \citealp{lan13}). The minimum and maximum values for $g_t$ have to be chosen sufficiently large to cover the essential domain of the log-volatility process, but not too large, in order to maintain sufficient fineness of the grid. More guidance on this issue is provided in \citet{lan12}. Another technical issue in the numerical maximization is that of local maxima: it may sometimes happen that the numerical search fails to find the MLE, and return a local maxima instead. The best way to address this issue seems to be to use a number of different sets of initial values in order to find and verify the global maximum. Uncertainty quantification, for both the parameters of the underlying log-volatility process and the density of $\varepsilon_t$, can be conducted using a parametric bootstrap, but the computational burden is relatively high.  
 
\subsubsection{Choice of the smoothing parameter}
\label{smoothchoice}

Cross-validation techniques can be used to choose the smoothing parameter. For a given series of log-returns, we suggest to generate $C$ random partitions such that in each partition a suitable percentage of the observations form the calibration sample, while the remaining observations constitute the validation sample. For each of the $C$ partitions and any given ${\lambda}$, the model is then calibrated by estimating the parameters using only the calibration sample (treating the data points from the validation sample as missing data). 
Subsequently, proper scoring rules \citep{gne07} can be used on the validation sample to assess the calibrated model for the given ${\lambda}$. For computational convenience, we consider the log-likelihood of the validation sample (now treating the data points from the calibration sample as missing data), under the model fitted in the calibration stage, as the score of interest. From some pre-specified set of possible smoothing parameters, e.g., $\{ 2^n | n=r,r+1,\ldots,s\}$ where $r$ and $s$ are integers, we then select the ${\lambda}$ with the highest mean score, over all $C$ cross-validation samples. The number of samples $C$ needs to be high enough to give meaningful scores, 
but must not be too high to allow for the approach to be computationally feasible. In our real data analyses (see below), the consideration of $C=40$ samples led to stable results yet was computationally feasible.

\subsubsection{Model checking}

We have already seen that the use of the HMM forward algorithm provides an efficient and convenient way to evaluate the SV model likelihood. HMM machinery can also be exploited in order to check the goodness of fit of a given model. Following \citet{zuc09}, we consider one-step-ahead forecast pseudo-residuals, which are given by
 \begin{linenomath*}
\begin{align*}
r_t=\Phi^{-1}\bigl( F(y_t \mid y_{t-1}, y_{t-2}, \ldots , y_1) \bigr) \, .
\end{align*}
 \end{linenomath*}
Here $\Phi$ denotes the cumulative distribution function of the standard normal distribution, and $F(y_t \mid y_{t-1}, y_{t-2}, \ldots , y_1)$ is the cumulative distribution function of $y_t$ given all observations up to time $t-1$. Using the HMM-type approximation, this can be written as
 \begin{linenomath*}
\begin{align}\label{cdf}
F(y_{t} \mid y_{t-1},y_{t-2}, \ldots , y_1) & \approx  \sum_{i=1}^m \zeta_i F(y_{t} \mid g_t = b_i^*) \\
\nonumber                                   & = \sum_{i=1}^m \zeta_i \int_{-\infty}^{y_t} \exp(-b_i^*/2) f_{\varepsilon} \left( x \exp(-b_i^*/2) \right) dx  \,,
\end{align}
 \end{linenomath*}
where $\zeta_i$ is the $i$th entry of the vector $\tilde{\boldsymbol{\alpha}}_{t-1} \boldsymbol{\Omega}/(\tilde{\boldsymbol{\alpha}}_{t-1}\mathbf{1'})$, which is defined as
 \begin{linenomath*}
\begin{equation*}
\tilde{\boldsymbol{\alpha}}_{t-1}=\boldsymbol{\delta}\mathbf{P}(y_1)\boldsymbol{\Omega}\mathbf{P}(y_2)\boldsymbol{\Omega}\cdots \boldsymbol{\Omega}\mathbf{P}(y_{t-1})\, ,
\end{equation*}
 \end{linenomath*}
$t=2,\ldots,T$, with $\boldsymbol{\delta}$, $\mathbf{P}(y_k)$ and $\boldsymbol{\Omega}$ defined as above. These $\tilde{\boldsymbol{\alpha}}_{t}$'s constitute the SV model analogue to the HMM forward probabilities; see the appendix for more details on the latter. The representation given in (\ref{cdf}) is only approximate due to the discretization of the log-volatility process, but as for the likelihood the accuracy also of this approximation can be made arbitrarily accurate by increasing $m$. In the context of SV models such residuals were first used by \citet{kim98}. If the fitted model is correct, then the pseudo-residuals are distributed standard normal. Thus, forecast pseudo-residuals can be used to identify extreme values, and the general suitability of the model can be checked by using, e.g., quantile-quantile-plots or formal tests for normality. 

\subsubsection{Decoding}

Again building on existing HMM machinery, estimates of the underlying log-volatility can easily be obtained using the Viterbi algorithm, which is an efficient dynamic programming algorithm for computing the most likely Markov chain state sequence to have given rise to observations stemming from an HMM (see \citealp{lan12}, or Chapter 5 of \citealp{zuc09}).

\section{Simulation experiments}
\label{simul}

\begin{figure}[!htb]
\begin{center}
{\includegraphics*[width=0.75\textwidth]{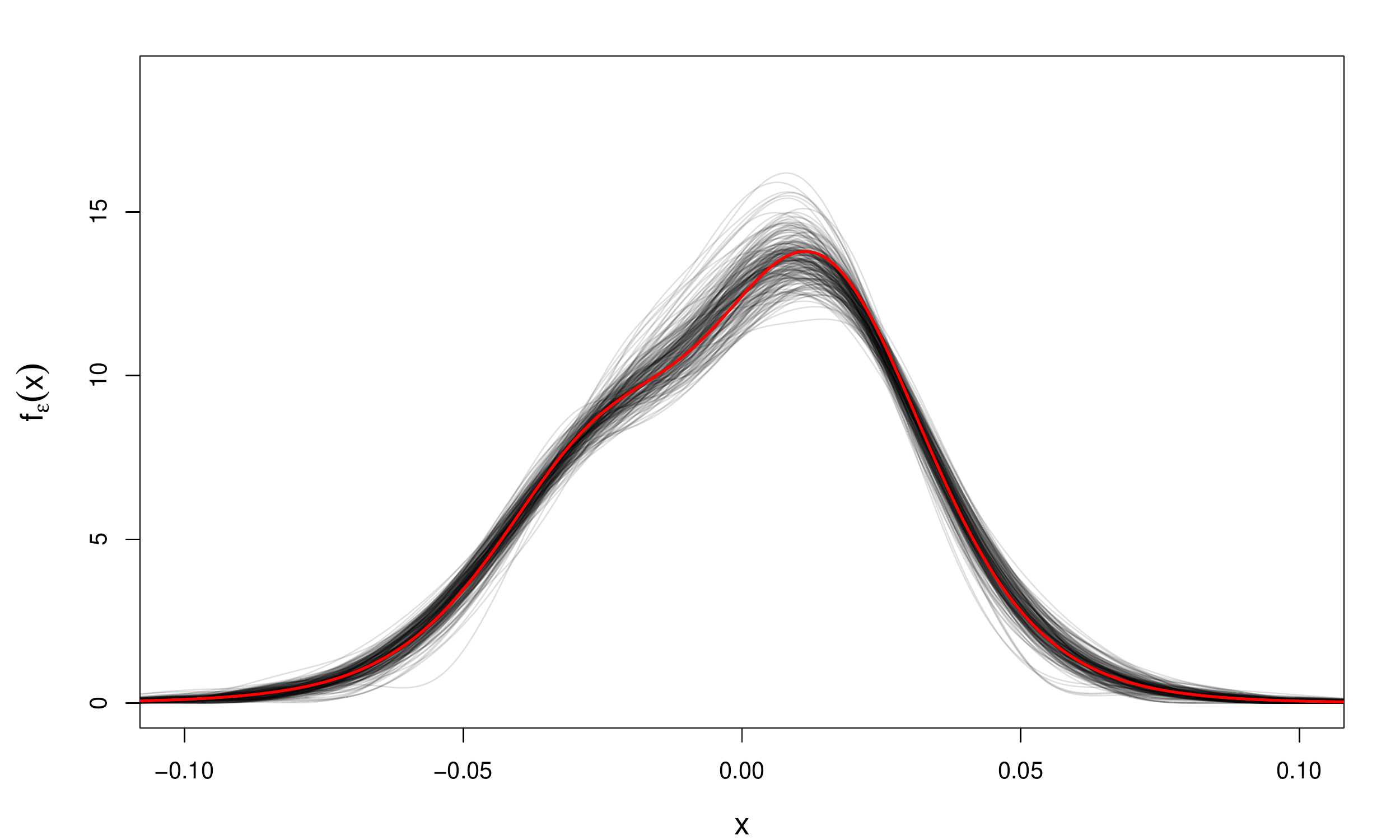}}
\end{center}
\vspace{-1em}
\caption{True density of $\varepsilon_t$ considered in the simulation experiments (red line) and its 200 estimates obtained using the nonparametric approach (grey lines).} \label{simres}
\end{figure}

To generate artificial data which adhere to many of the stylized facts discussed above, we use an SV model as in (\ref{SVnp}), with $\phi=0.98$, $\sigma=0.1$ and $\varepsilon_t$ specified as $\varepsilon_t= 0.02 (\zeta_{1,t}-1)^{\alpha_t} (\zeta_{2,t}+1)^{1-\alpha_t}+ 0.006$, where $\zeta_{1,t}$ and $\zeta_{2,t}$ are mutually independent iid sequences of Student-t random variables with 6 and 8 degrees of freedom, respectively, and $\alpha_t$ are iid Bernoulli variables, each taking on the value 1 with probability 0.35. This specification results in a skewed and leptokurtic distribution (skewness $\approx -0.22$; kurtosis $\approx 3.58$) with zero mean. For an illustration of the shape of this distribution, see Figure \ref{simres}.

For this model, we conducted 200 simulation runs, with $T=4000$ observations being generated in each run. In each run, the final 1000 observations of the generated series were used only to assess the predictive capacity of various models, which were previously fitted to the first 3000 observations. To make a fairly extensive simulation study feasible, we did not conduct a cross-validation for the smoothing parameter $\lambda$ within each simulation run. Instead, we ran cross-validations only in 10 preliminary simulation runs, trying the values $256$, $512$, $1024$, $2048$, $4096$ and $8192$, and then fixed $\lambda$ for the main 200 simulation runs at the value which was selected most often by cross-validation in the preliminary runs (namely $\lambda=1024$). This procedure resulted in a good performance (see below), but in fact the results could potentially be further improved by conducting a cross-validation within each simulation run. We set $K=15$, resulting in 31 B-spline basis densities that were used in the estimation. To obtain a benchmark for the semiparametric model SV$_{\text{sp}}$, we further fitted the basic models SV$_0$ and SV$_{\text{t}}$ to each generated series, also using the HMM-based discretization approach described in Section \ref{modelsubsec}. 

For the SV$_{\text{sp}}$ model, the sample mean estimates of the parameters $\phi$ and $\sigma$ were obtained as $0.978$ (sample standard error: $0.007$) and $0.103$ ($0.017$), respectively. For the SV$_{0}$/SV$_{\text{t}}$ model, the sample mean estimates of the parameters $\phi$ and $\sigma$ were obtained as $0.972$/$0.977$ (sample standard errors: $0.010$/$0.007$) and $0.117$/$0.102$ ($0.023$/$0.017$), respectively. The SV$_{0}$ model underestimated the persistence parameter and overestimated the variance of the log-volatility process. The latter is due to the model's inability to capture the slight excess kurtosis of the true conditional distribution (see further comments on this issue in Section \ref{appl2}). In contrast, both the SV$_{\text{sp}}$ model and the SV$_{\text{t}}$ model yielded approximately unbiased estimates of the parameters related to the log-volatility process. Concerning the conditional process, $y_t$, Figure \ref{simres} displays the true pdf of $\varepsilon_t$ and the corresponding pdfs that were estimated using the nonparametric approach. From the graphic we can see that all 200 fits seem fairly reasonable.

In the given scenario, we also assessed the predictive capacity of the three different modelling approaches, represented by the models SV$_0$, SV$_{\text{t}}$ and SV$_{\text{sp}}$. We did this by calculating, in each simulation run and under each of the three types of models fitted to the first 3000 observations, the log-likelihood score for the final 1000 observations, denoted by $\text{llk}_i\text{(SV}_{0})$, $\text{llk}_i\text{(SV}_{\text{t}})$ and $\text{llk}_i\text{(SV}_{\text{sp}})$, with $i$ indicating the simulation run. These scores were compared to the corresponding score obtained when using the true model, i.e., the one that was actually used to generate the artificial data; we denote this score by $\text{llk}_i\text{(SV}_{\text{true}})$. In this simulation experiment, with skewed and leptokurtic conditional distribution, the averaged differences between the scores obtained for SV$_0$, SV$_{\text{t}}$ and SV$_{\text{sp}}$ on the one hand, and the score obtained for the true model on the other hand, were obtained as
 \begin{linenomath*}
\begin{align*}
\frac{1}{200}\sum_{i=1}^{200} \bigl( \text{llk}_i\text{(SV}_{\text{0}})   - \text{llk}_i\text{(SV}_{\text{true}}) \bigl)  & = -9.48 \, , \\
\frac{1}{200}\sum_{i=1}^{200} \bigl( \text{llk}_i\text{(SV}_{\text{t}})   - \text{llk}_i\text{(SV}_{\text{true}}) \bigl)  & = -9.14 \quad \text{and} \\
\frac{1}{200}\sum_{i=1}^{200} \bigl( \text{llk}_i\text{(SV}_{\text{sp}})  - \text{llk}_i\text{(SV}_{\text{true}}) \bigl)  & = -3.52 .
\end{align*}
 \end{linenomath*}
On average, the SV$_{\text{sp}}$ model hence fitted the out-of-sample data substantially better than its parametric counterparts. Considering the individual simulation runs, the SV$_{\text{sp}}$ model had a better predictive performance than the two parametric models in 176 out of 200 cases. These results are hardly surprising given the skewness of the distribution chosen for $\varepsilon_t$. Nevertheless, they do demonstrate both the practical feasibility and the potential benefits of our approach. 

In order to have a benchmark for these results, we ran a second simulation study in which we generated data from the SV$_{\text{t}}$ model, specifying $\varepsilon_t^{(0)}$ in (\ref{SVbasic}) to be a Student-t distribution with 10 degrees of freedom and $\mu=0.02$. Except of the true distribution of the conditional distribution, this second simulation experiment was configured exactly as the first. Again considering the three models SV$_0$, SV$_{\text{t}}$ and SV$_{\text{sp}}$, we obtained
 \begin{linenomath*}
\begin{align*}
                 \frac{1}{200} \sum_{i=1}^{200} \bigl( \text{llk}_i\text{(SV}_{\text{0}})   - \text{llk}_i\text{(SV}_{\text{true}}) \bigl)  & = -5.44 \, , \\
\displaybreak[1] \frac{1}{200} \sum_{i=1}^{200} \bigl( \text{llk}_i\text{(SV}_{\text{t}})   - \text{llk}_i\text{(SV}_{\text{true}}) \bigl)  & = -0.62 \quad \text{and} \\
                 \frac{1}{200} \sum_{i=1}^{200} \bigl( \text{llk}_i\text{(SV}_{\text{sp}})  - \text{llk}_i\text{(SV}_{\text{true}}) \bigl)  & = -2.30 \, .
\end{align*}
 \end{linenomath*}
In this scenario, the (correct) SV$_{\text{t}}$ model had a better predictive performance than the two other models in 152 out of 200 cases, while the SV$_{\text{sp}}$ model still showed a better predictive performance than the (incorrect) SV$_{0}$ model in 164 cases. Thus, overall, the results demonstrate a) the potential of the nonparametric approach to considerably improve the predictive capacity in scenarios where the true conditional distribution deviates from the functional form imposed by either the normal or the Student-t distribution (e.g., if it is skewed), and b) that it can perform almost as well as parametric modelling approaches in scenarios where those are adequate.

\section{Application to stock returns}\label{appl}

\subsection{The data}\label{appl1}

\begin{figure}[!htb]
\begin{center}
{\includegraphics*[width=0.9\textwidth]{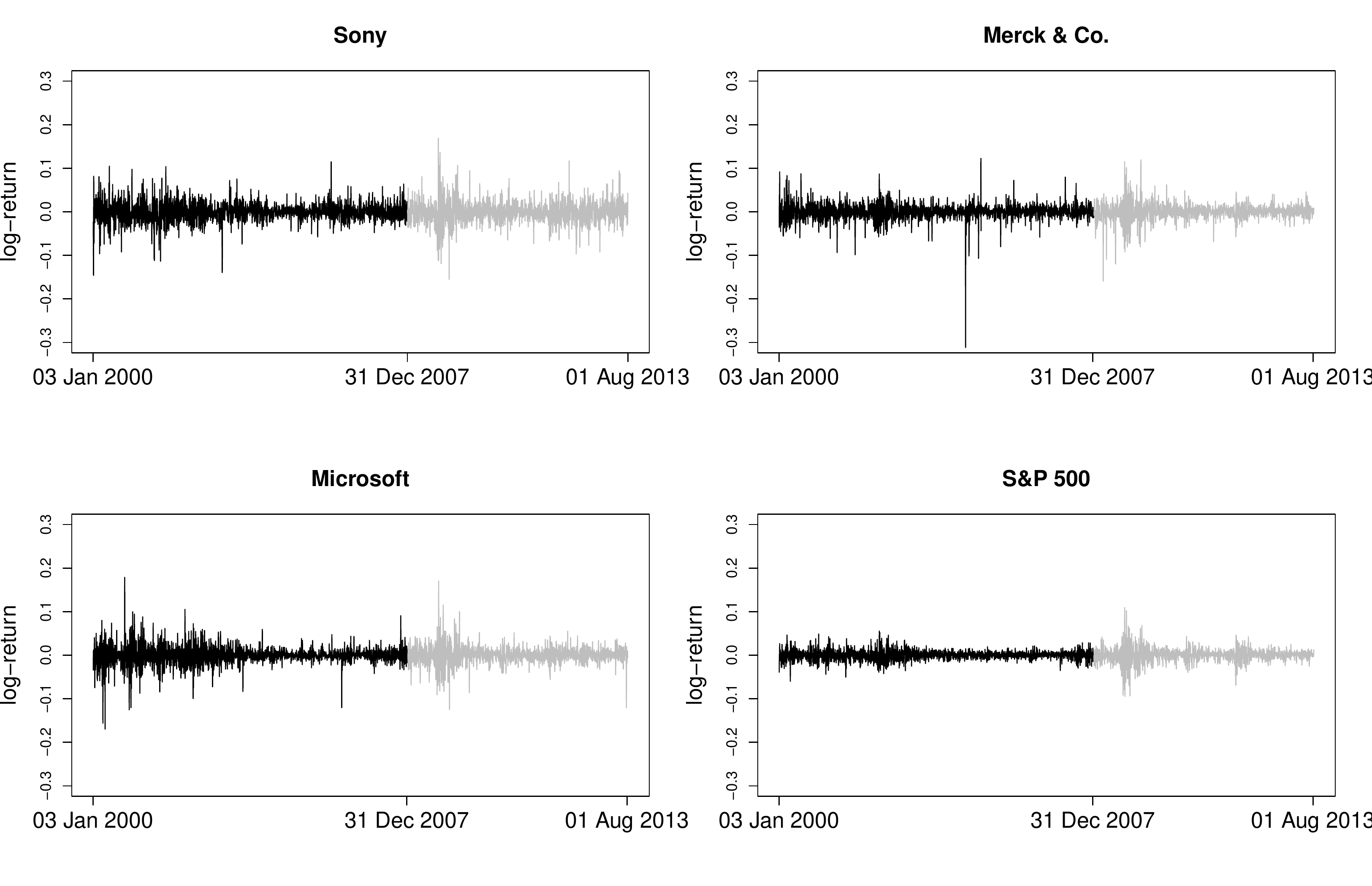}}
\end{center}
\caption{Time series of log-returns on shares of three companies (Sony, Merck and Microsoft) and on the stock index S{\&}P 500; the observations from the in-sample and out-of-sample periods are displayed in black and gray, respectively.} \label{realdata}
\end{figure}

The SV$_{\text{sp}}$ model was fitted to series of daily log-returns for three stocks, namely Sony Corporation, Merck \& Co.\ and Microsoft Corporation, and for the stock index S{\&}P 500.
The adjusted closing prices, $p_t$, for the period 03.01.2000\,--\,01.08.2013, were downloaded from ``finance.yahoo.com'', and the daily log-returns were computed as $y_t = \log(p_t/p_{t-1})$.
To assess the out-of-sample predictive performance of various models, we divided each of the four series into two parts: 
\begin{itemize}
\item In-sample period: 03.01.2000\,--\,31.12.2007, 
\item Out-of-sample period: 02.01.2008\,--\,01.08.2013. 
\end{itemize}
The dividing date was chosen to lie before the outburst of the recent financial crisis, which culminated in the collapse of Lehman Brothers Holdings Inc.\ in September 2008. The four time series that were analyzed are displayed in Figure \ref{realdata}. 

For comparison purposes, we also fitted the two basic models SV$_0$ and SV$_{\text{t}}$ to each of the four series, using the same partition into in-sample period and out-of-sample period. All models were fitted using the data from the in-sample period only, and the data from the out-of-sample period were used to assess the predictive performance of the three different models (as detailed below).

\subsection{Results}\label{appl2}

To each of the four series, the SV$_{\text{sp}}$ model was fitted using $K=20$ and hence 41 B-spline basis densities to represent the density of $\varepsilon_t$, $m=100$ intervals in the discretization of the log-volatility process, numerically integrating over the log-volatility values from the interval $[b_0,b_{100}]$=$[-5,5]$. 
Smoothing parameters were selected via cross-validation as described in Section \ref{smoothchoice}, in each of $C=40$ cross-validation partitions using 90\% of the data points in the calibration stage. 
Fitting the SV$_{\text{sp}}$ model took about 10 minutes per series on an i7 CPU, at 2.7 GHz and with 4 GB RAM.

\begin{table}[!htb]
\setlength\columnsep{-2em}
\caption{Parameter estimates that were obtained when fitting the three different models to the four series of log-returns considered.}
\label{paraest}
\begin{center}
\begin{tabular}{l|ccc|cccc|ccccccccccc}
               & SV$_0$ & &                                         &  SV$_{\text{t}}$ & & &                                            & SV$_{\text{sp}}$ & \\
               & $\hat{\phi}$  & $\hat{\sigma}$  & $\hat{\beta}$    &  $\hat{\phi}$  & $\hat{\sigma}$  & $\hat{\beta}$ & $\hat{\nu}$    &  $\hat{\phi}$  & $\hat{\sigma}$       \\[0.1em]
\hline
Sony           & 0.957         & 0.249           & 0.019            &  0.992         & 0.092           & 0.017         & 6.698          &  0.994         & 0.087                \\
Merck          & 0.825         & 0.545           & 0.014            &  0.992         & 0.086           & 0.012         & 4.670          &  0.993         & 0.078                \\
Microsoft      & 0.979         & 0.239           & 0.015            &  0.994         & 0.116           & 0.014         & 6.305          &  0.994         & 0.122                \\
S{\&}P 500     & 0.991         & 0.114           & 0.010            &  0.992         & 0.104           & 0.009         & 25.724         &  0.998         & 0.088
\end{tabular}
\end{center}
\end{table}
\setlength\columnsep{+1em}

The estimates of the parameters $\phi$, $\sigma$, $\beta$ (only for the SV$_0$ and for the SV$_{\text{t}}$ model) and $\nu$ (only in case of the SV$_{\text{t}}$ model), for the four series considered, are given in Table \ref{paraest}. The results obtained for the SV$_0$ model illustrate the problems of the conditional normal distribution to capture the extreme returns. Especially for the Merck stock, where on September 30, 2004, the withdrawal of the drug Rofecoxib from the market caused heavy losses, SV$_0$ performs badly. Indeed, the only way for the SV$_0$ model to cope with the associated extreme negative return of $-0.31$, which occurs in a period of calm market, is to assign a very high uncertainty to the log-volatility process (as expressed by a high $\hat{\sigma}$ and a small $\hat{\phi}$). This results in an undersmoothing of the volatility. By contrast, the SV$_{\text{t}}$ model's leptokurtic conditional distribution leads to much more plausible estimates for $\phi$ and $\sigma$, with the results being similar to those obtained for the SV$_{\text{sp}}$ model. In these two models, extreme returns are assigned to the tail of the return distribution, rather than to big jumps in the log-volatility process, as in the SV$_0$ model. The same pattern is found for the other two stock return series --- although to a lesser extent. Only for the stock index S{\&}P 500, the estimate of $\sigma$ obtained using the SV$_0$ model is of the same magnitude as the corresponding estimates obtained when applying SV$_{\text{t}}$ and SV$_{\text{sp}}$. This is not surprising since in the index the extreme returns of individual companies play a smaller role, which is also reflected by the much lighter tail of the conditional distribution in the fitted SV$_{\text{t}}$ model.

\begin{figure}[!htb]
\begin{center}
{\includegraphics*[width=0.95\textwidth]{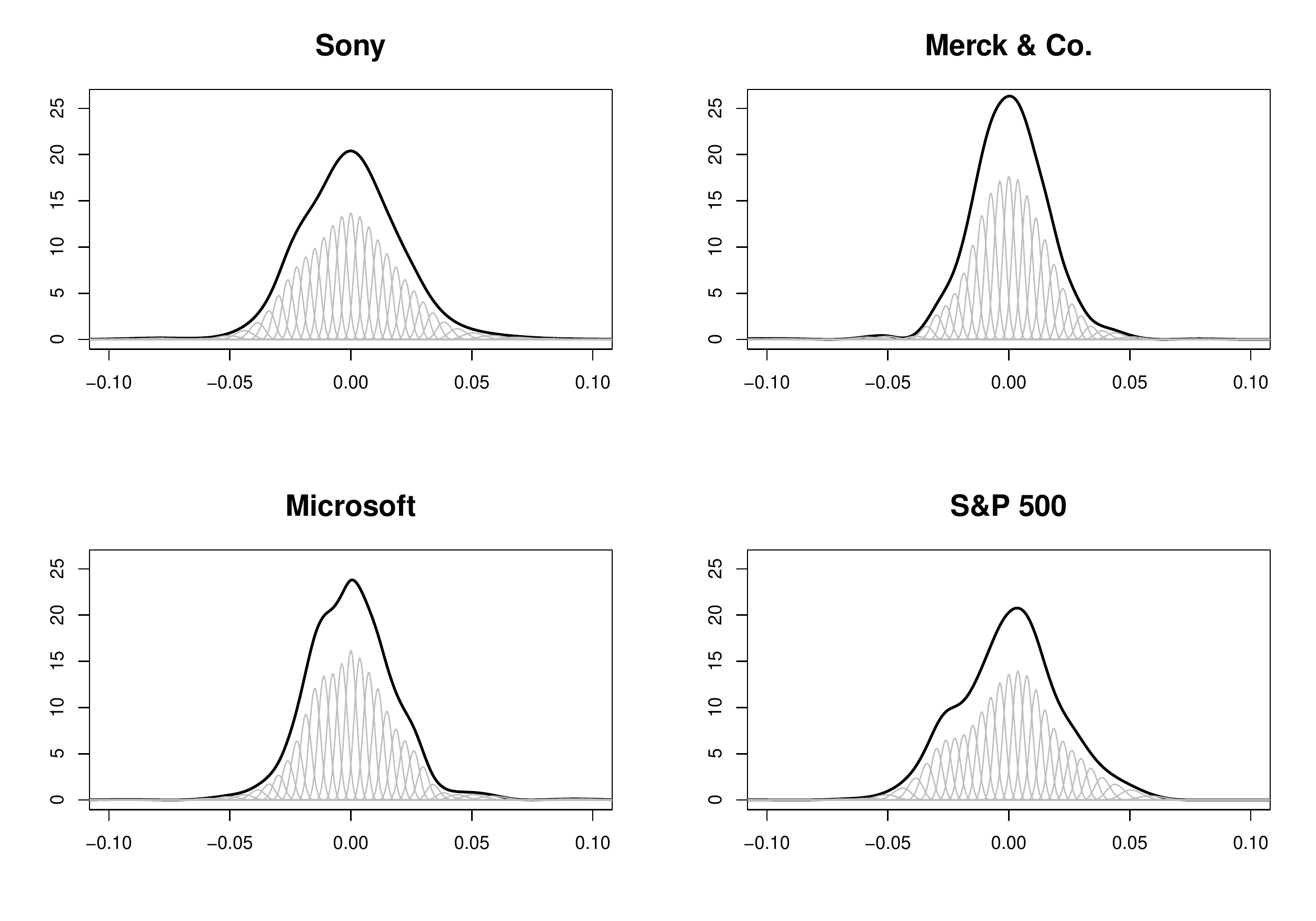}}
\end{center}
\caption{Conditional densities of $\varepsilon_t$ estimated using the nonparametric approach, for the four series of log-returns considered, and underlying weighted B-splines that generate these densities via a linear combination (in grey).} \label{fitteddist}
\end{figure}

Figure \ref{fitteddist} displays the nonparametrically estimated densities of the conditional distribution in the SV$_{\text{sp}}$ model, for the four series considered. The skewness of the nonparametrically estimated distribution $f_{\varepsilon}$ is $-0.73$, $-2.61$, $-1.23$ and $-0.77$, for Sony, Merck, Microsoft and S{\&}P 500, respectively. This is in line with the stylized facts
attributed to financial assets which propagate a gain/loss asymmetry as large drawdowns generally exceed large upward movements \citep{con01}, resulting in a left tail in the conditional distribution that is more extreme than the right tail \citep{dur06}. However, to some extent the skewness also stems from an asymmetric density close to the center, a phenomenon that could be related to the lack of a leverage effect in our model. Indeed, \citet{del13} gave evidence that if data stem from an SV model with a strong leverage effect, then a nonparametric SV model not including this effect may infer a multimodal conditional distribution. 

\begin{figure}[!htb]
\begin{center}
{\includegraphics*[width=1\textwidth]{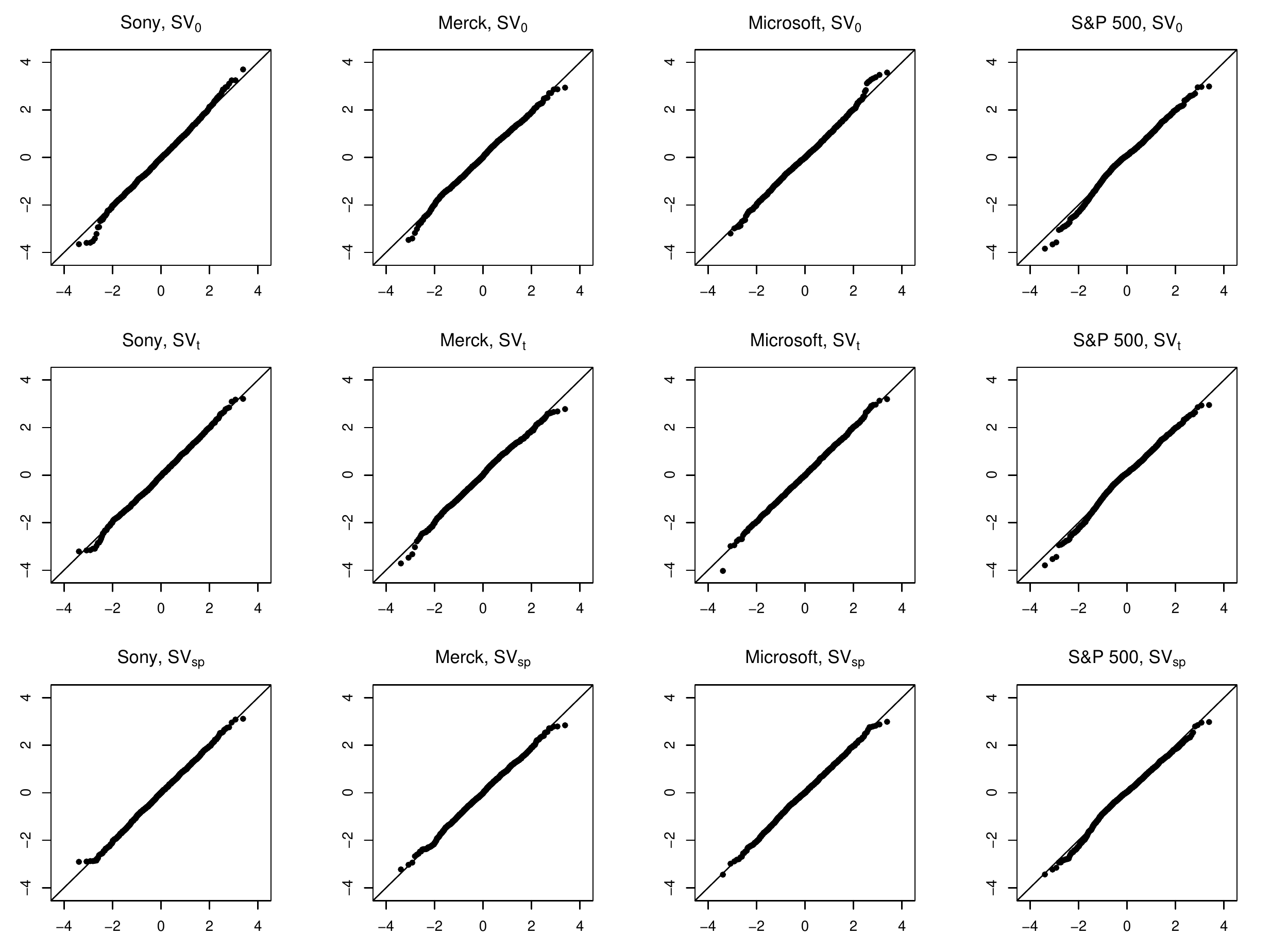}}
\end{center}
\vspace{-1em}
\caption{Assessment of the absolute fit of the models in terms of their predictive performance: quantile-quantile plots of the out-of-sample forecast pseudo-residuals obtained from the fitted models SV$_0$ (top row), SV$_{\text{t}}$ (second row) and SV$_{\text{sp}}$ (bottom row), for Sony (first column), Merck (second column), Microsoft (third column) and S{\&}P 500 (fourth column). Sample quantiles are given on the vertical axes, and quantiles of the standard normal are given on the horizontal axis.} \label{qq}
\end{figure}

Quantile-quantile plots of the out-of-sample one-step-ahead forecast pseudo-residuals associated with the returns observed in the out-of-sample period 02.01.2008\,--\,01.08.2013, under the models fitted to the data from the in-sample period, are given in Figure \ref{qq}. The p-values for the Jarque--Bera tests applied to the pseudo-residuals are listed in Table \ref{jbt}. The results show that the SV$_0$ model provides poor out-of-sample forecasts for all three stock price time series, as the corresponding pseudo-residuals show large deviations from normality. The SV$_{\text{t}}$ model is able to adequately forecast the Sony and Microsoft stocks, but exhibits some problems in the forecasts for the Merck series. Overall, the SV$_{\text{sp}}$ shows a slightly higher accuracy in the forecasts of the stock returns, especially with respect to the extreme negative returns. For the S{\&}P 500 index, all three models perform badly, with the Jarque-Bera test rejecting the null hypothesis of normally distributed pseudo-residuals in each case. While for the models SV$_0$ and SV$_{\text{t}}$ at least part of the reason for the poor performance is the inaccurate forecast of extreme negative returns, for the SV$_{\text{sp}}$ model the reason for the poor performance lies solely in the inaccurate forecast of moderate losses, while more extreme losses are again captured well. We believe the reason for the bad performance in this particular case to lie in the long persistent decline of stock prices during the financial and economic crisis which ensued the collapse of Lehman Brothers.

\begin{table}[!htb]
\caption{Assessment of the absolute fit of the models in terms of their predictive performance: p-values of Jarque--Bera tests applied to out-of-sample one-step-ahead ahead forecast pseudo-residuals.}
\label{jbt}
\begin{center}
\begin{tabular}{l|llllllllllll}
                  & Sony                                          & Merck                                   & Microsoft                    & S{\&}P 500                    \\[0.1em]
\hline
SV$_0$            & $0.048^{**}$ & $0.002^{**}$ & $<0.001^{**}$ & $<0.001^{**}$  \\
SV$_{\text{t}}$   & $0.896$ & $0.055^*$ & $0.795$ & $<0.001^{**}$  \\
SV$_{\text{sp}}$  & $0.818$ & $0.333$ & $0.407$ & $<0.001^{**}$  
\end{tabular}
\end{center}
\end{table}

As in the simulation study, we also calculated the log-likelihood scores for the observations from the out-of-sample period, for the three different models and each of the four series analyzed. The results of this comparative assessment of the out-of-sample predictive performance of the models are displayed in Table \ref{llks}. The log-likelihood scores portray a similar picture as above, with the SV$_0$ model performing worse than the two models SV$_{\text{t}}$ and SV$_{\text{sp}}$ (at least for all stock returns), which again perform similarly well. While the SV$_{\text{sp}}$ model performs slightly better for the Sony and Merck time series, SV$_{\text{t}}$ has the edge for Microsoft and S{\&}P 500. 


With regard to the index S{\&}P 500, as well as other indices, it should be noted that these exhibit less extreme dynamics due to the averaging over several stocks. As a consequence, the simple model SV$_0$ in this case performs about as well as do the more flexible models SV$_{\text{t}}$ and SV$_{\text{sp}}$. However, the quantile-quantile plots do indicate that even for the index S{\&}P 500 our semiparametric model has a slightly improved forecast accuracy at the extreme end of the lower tail. 

\begin{table}[!htb]
\caption{Assessment of the relative fit of the models in terms of their predictive performance: log-likelihood scores for the out-of-sample period. For each company, the highest score is underlined.}
\label{llks}
\begin{center}
\begin{tabular}{l|lllllllll}
                                     & Sony    & Merck     & Microsoft     & S{\&}P 500     \\[0.1em]
\hline
$\text{llk}\text{(SV}_{0})$          & 3265.31   & 3891.75     & 3778.64         & 4228.95 \\
$\text{llk}\text{(SV}_{\text{t}})$   & 3269.58   & 3913.12     & \underline{3799.58}         & \underline{4230.53} \\
$\text{llk}\text{(SV}_{\text{sp}})$  & \underline{3271.38}   & \underline{3914.65}     & 3798.64         & 4228.74  
\end{tabular}
\end{center}
\end{table}

\section{Discussion}\label{discuss}

The stylized facts of asset returns indicate that simple parametric distributions, such as the normal or the Student's t-distribution, may not be well-suited to describe the shape of the conditional distribution in SV models. Thus, a nonparametric modelling of the conditional distribution, which allows for heavy tails, gain/loss asymmetry and other unusual features, may bear considerable advantages. In this manuscript, we developed a powerful and flexible frequentist framework for a nonparametric estimation of the conditional distribution in a discrete-time SV model. The approach exploits the strengths of the HMM machinery, in particular allowing for model checking, forecasting and volatility estimation. 

The computational burden for estimating a model of the proposed type is low, namely in the order of a couple of minutes for the considered series and fixed smoothing parameter. Applying cross-validation techniques to choose a data-driven smoothing parameter is, however, computationally demanding: computing times for this part of the analysis were about 10-15 hours per series we analyzed. These computing times can be substantially reduced by employing parallel computing, allowing for computing times below one hour. Although the model specifications are not directly comparable, \citet{del11} report computing times of up to a day for the model fitting with their Bayesian approach, for series that were slightly longer than the ones we considered and when fixing the concentration parameter for the Dirichlet process mixture (which represents an analogue to the smoothing parameter in our setting). 

A technical issue with the presented method which calls for further research concerns the configuration of the B-spline basis densities used in the estimation. We employed an ad hoc approach to account for the fact that in the tails of the conditional distribution only few observations are available to infer the shape of the density. Our approach effectively increases the penalty for non-smoothness in the tails of the distribution. The use of equally spaced sample quantiles, as suggested by \citet{rup02}, seems a promising avenue to explore in this regard. An alternative would be to follow the literature on adaptive smoothing parameter selection, e.g., \citet{kri08}, where the smoothing parameter would be specified as another spline function on the log-volatility domain. However, this would considerably increase the complexity of the likelihood-based analysis and may therefore easily lead to numerical or identifiability problems. 

While we modelled the conditional distribution in the SV model in a nonparametric way, we still assumed a parametric distribution form of the innovations in the log-volatility process, which is not necessary. Furthermore, the possible incorporation of leverage effects into the model --- i.e., the explicit modelling of a (negative) correlation between returns and subsequent log-volatilities, as often done in parametric SV modelling (e.g., \citealp{har96}, \citealp{jac04}) --- was not discussed in the present manuscript, since we felt that in this first step towards a frequentist framework for semiparametric SV modelling it is advisable to focus on the inferential machinery, rather than on exploring the various possible variations in the model structure. Corresponding extensions are to be explored in future research.  

Overall, the approach shows promise as a useful novel tool for analyzing time series of daily log-returns. We have illustrated that the approach {\em can} lead to an improved predictive capacity compared to basic parametric SV models, and in the real data analyses we found some notable distributional shapes. In particular, our model revealed negative skewness and heavy tails in the conditional distribution of the returns we analyzed, while still identifying the behavior of the log-volatilities that is typical of SV models. In out-of-sample comparisons the parametric model with Student-t conditional distribution performed similarly compared to our semiparametric model (and both performed much better than the model with Gaussian conditional distribution, at least for stock returns). It should be noted here that all validation samples that we considered involve rather extreme dynamics as they comprise the recent financial crisis. Further research needs to be done to investigate the performance of our approach in different scenarios, including calmer markets. The present manuscript is intended mainly to introduce the frequentist estimation framework and to outline its potential. 
While much work remains to be done, we strongly believe that the model's flexibility with regard to describing asymmetries and extreme events, and the relative accessibility of the maximum likelihood framework, will render our approach and potential future extensions a useful tool in portfolio management.



\section*{Supplementary Material}
R and C++ code to 1) generate artificial data as in the simulation study and to 2) fit, to the generated data, the SV$_{0}$, SV$_{\text{t}}$ and SV$_{\text{sp}}$ models (format: 
main R code in .R file and additional C++ code in .cpp file).

\section*{Appendix -- HMM essentials}

This appendix reviews some HMM basics. A standard $m$-state HMM has the same two process structure as SV models and SSMs, only that the unobserved process is a Markov chain and hence discrete- rather than continuous-valued. Consider an HMM with observable process $\{ X_t \}_{t=1}^T$ and underlying Markov chain $\{ S_t \}_{t=1}^T$. Given the current state of $S_t$, the variable $X_t$ is usually assumed to be conditionally independent from previous and future observations and states. The Markov chain is typically considered to be of first order, and the probabilities of transitions between the different states are summarized in the $m \times m$ transition probability matrix $\boldsymbol{\Gamma}=\left( \gamma_{ij} \right)$, where
$\gamma_{ij}=\Pr \bigl(S_{t+1}=j\vert S_t=i \bigr)$, $i,j=1,\ldots,m$.
The initial state probabilities are summarized in the vector $\boldsymbol{\pi}$, where $\pi_{i} = \Pr (S_1=i)$, $i=1,\ldots,m$. It is usually convenient and appropriate to assume $\boldsymbol{\pi}$ to be the stationary distribution. 
For the described HMM, with observations given by $x_1,\ldots,x_T$ and underlying states denoted by $s_1,\ldots,s_T$, the likelihood is given by
\begin{align*}
\mathcal{L}^{\text{HMM}}          = f(x_1, \ldots, x_T) 
                                 & = \sum_{s_1=1}^m \ldots \sum_{s_T=1}^m f(x_1, \ldots, x_T | s_1, \ldots, s_T) f(s_1, \ldots, s_T) \\
                                 & = \sum_{s_1=1}^m \ldots \sum_{s_T=1}^m \pi_{s_1} \prod_{t=1}^T f (x_t | s_t) \prod_{t=2}^T \gamma_{s_{t-1},s_t} \, .
\end{align*}
In this form the likelihood involves $m^T$ summands, which would make a numerical maximization infeasible in most cases. However, there is a much more efficient way of calculating the likelihood $\mathcal{L}^{\text{HMM}}$, given by a recursive scheme called the {\em forward algorithm}. To see this, we consider the vectors of forward variables, defined as
$\boldsymbol{\alpha}_t = \bigl( {\alpha}_t (1), \ldots , {\alpha}_t (m) \bigr)$, $t=1,\ldots,T$,
where ${\alpha}_t (j) = f (x_1, \ldots, x_t, S_t=j)$, $j=1,\ldots,m$. We then have the recursion:
 \begin{equation}\label{forw}
 \boldsymbol{\alpha}_1  = \boldsymbol{\pi} \mathbf{Q}(x_1) \, , \qquad  \boldsymbol{\alpha}_{t+1}  =  \boldsymbol{\alpha}_{t} \boldsymbol{\Gamma} \mathbf{Q}(x_{t+1}) \, ,
 \end{equation}
where $\mathbf{Q}(x_t)= \text{diag} \bigl( f_1 (x_{t}), \ldots, f_m (x_{t}) \big)$, with $f_i(x_t) =f (x_{t} | S_{t}=i)$. The recursion (\ref{forw}) can be derived in a straightforward manner using the HMM dependence structure. The likelihood can then be written as a matrix product:
\begin{equation*}\label{lik}
\mathcal{L}^{\text{HMM}} = \sum_{i=1}^m {\alpha}_T(i) = \boldsymbol{\pi} \mathbf{Q}(x_1) \boldsymbol{\Gamma} \mathbf{Q}(x_{2}) \ldots \boldsymbol{\Gamma} \mathbf{Q}(x_{T}) \mathbf{1} \, ,
\end{equation*}
where $\mathbf{1}\in \mathbbm{R}^m$ is a column vector of ones. For more details on HMMs, see, e.g., \citet{zuc09}.

\end{spacing}

\end{document}